\documentclass[twocolumn,showpacs,superscriptaddress,prl,amsmath,amssymb,letterpaper]{revtex4}
\usepackage{times}
\usepackage{amsmath,bm,amsfonts}
\usepackage{dcolumn}
\usepackage{graphicx}
\usepackage{latexsym}

\begin{document}

\title{Theory of topological quantum phase transitions in 3D noncentrosymmetric systems}

\author{Bohm-Jung \surname{Yang}}
\affiliation{RIKEN Center for Emergent Matter Science, Wako, Saitama 351-0198, Japan}

\author{Mohammad Saeed \surname{Bahramy}}
\affiliation{RIKEN Center for Emergent Matter Science, Wako, Saitama 351-0198, Japan}

\author{Ryotaro \surname{Arita}}
\affiliation{RIKEN Center for Emergent Matter Science, Wako, Saitama 351-0198, Japan}

\affiliation{Department of Applied Physics, University of Tokyo,
Tokyo 113-8656, Japan}

\author{Hiroki \surname{Isobe}}
\affiliation{Department of Applied Physics, University of Tokyo,
Tokyo 113-8656, Japan}

\author{Eun-Gook \surname{Moon}}
\affiliation{Department of Physics, University of California,
Santa Barbara, CA 93106, USA}

\author{Naoto \surname{Nagaosa}}
\affiliation{RIKEN Center for Emergent Matter Science, Wako, Saitama 351-0198, Japan}

\affiliation{Department of Applied Physics, University of Tokyo,
Tokyo 113-8656, Japan}

\date{\today}

\begin{abstract}
We have constructed a general theory describing the topological quantum phase transitions in 3D systems with broken inversion symmetry.
While the consideration of the system's codimension generally predicts the appearance of
a stable metallic phase
between the normal and topological insulators,
it is shown that a direct topological phase transition between two insulators is also possible
when an accidental band crossing (ABC) occurs along directions with high crystalline symmetry.
At the quantum critical point (QCP), the energy dispersion becomes quadratic along
one direction while the dispersions along the other two orthogonal directions are linear,
which manifests the zero chirality of the band touching point (BTP).
Due to the anisotropic dispersion at QCP, various thermodynamic and transport
properties show unusual temperature dependence and anisotropic behaviors.
\end{abstract}


\maketitle

The 3D topological insulator (TI) is a new state of matter
in which the nontrivial topology of bulk electronic wave functions
guarantees the existence of gapless states on the sample's boundary.~\cite{Fu-Kane-Mele, Qi}
Because of its topological nature,
the surface gapless states
are protected against small perturbations preserving the time-reversal symmetry (TRS) as long as
the bulk band gap remains finite. Therefore to change
the bulk topological property, the band gap should be
closed at some points in the Brillouin zone (BZ) via accidental band crossing (ABC).
Recently, such a topological phase transition (PT) is realized
in BiTl(S$_{1-x}$Se$_{x}$)$_{2}$~\cite{QCP_exp1,QCP_exp2},
by modulating the spin-orbit interaction or
the crystal lattice. In inversion symmetric systems such as BiTl(S$_{1-x}$Se$_{x}$)$_{2}$,
the topological PT can be described by the (3+1) dimensional massive
Dirac Hamiltonian in general. In this sense, the topological PT
of 3D TIs provides a new venue to study intriguing quantum critical behaviors of 3D particles
with relativistic dispersion.~\cite{Chakravarty, Isobe, Hosur}

\begin{figure}[t]
\centering
\includegraphics[width=3 in]{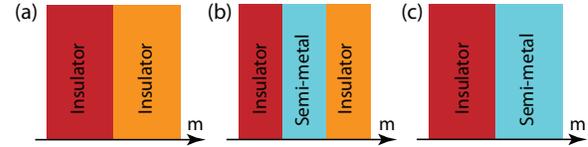}
\caption{(Color online)
Generic phase diagrams, resulting from ABC
between  conduction and valence bands in 3D noncentrosymmetric systems.
Here $m$ indicates an external control parameter.
} \label{fig:phasediagram}
\end{figure}

On the other hand, for noncentrosymmetric systems,
our understanding of the topological PT and
corresponding quantum critical behavior is still incomplete. By considering
the codimension for ABC,
a stable metallic phase was predicted to appear between TI
and normal insulator in 3D noncentrosymmetric systems.~\cite{Murakami}
The intervening metallic phase, dubbed a Weyl semi-metal,
has topological stability because there are several gapless points (Weyl points)
with nonzero chiral charge at the Fermi level.~\cite{Weyl_iridate} Therefore
before every Weyl point is annihilated by colliding with another Weyl point
with opposite chiral charge, the Weyl semi-metal should stably survive
across the PT.
In this respect, the recent discovery of a direct PT
between two insulators in noncentrosymmetric compound BiTeI is an unexpected surprise.~\cite{BiTeI_exp, BiTeI_theory1,BiTeI_theory2}
At the QCP of BiTeI, instead of a Weyl semi-metal, several isolated band touching points (BTPs) with anisotropic dispersion
appear, which suggests
the diversity of the possible phase diagrams of noncentrosymmetric systems
accessible via ABC.

In this paper, we propose generic phase diagrams for
3D noncentrosymmetric systems that can be achieved through
ABC, as depicted in Fig.~\ref{fig:phasediagram}.
We carry out the analysis of the minimal two-band Hamiltonian
describing the ABC
to derive the conditions for these insulator-to-metal and insulator-to-insulator transitions (IIT).
The key ingredient to obtain Fig.~\ref{fig:phasediagram} is the fact that
the chirality of the BTP at QCP is zero. Therefore it can be either gapped out
leading to another insulator (Fig.~\ref{fig:phasediagram} (a)) or
split into several Weyl points resulting in a Weyl semi-metal.
In the latter case, depending on whether the trajectory, traversed by
the Weyl point, is closed or not,
the Weyl semi-metal phase turns into another insulator (Fig.~\ref{fig:phasediagram} (b))
or persists all the way (Fig.~\ref{fig:phasediagram} (c)).
In all three cases, at the QCP between any pair of neighboring phases,
the energy dispersion near a BTP is highly anisotropic, which is
linear in two directions
and quadratic along the third direction.
This anisotropic dispersion induces new power laws
in the temperature dependence of various measurable quantities
and anisotropic physical responses.

\textit{Phase transition through ABC.$-$}
In noncentrosymmetric systems,
the ABC between the conduction and valence bands
can be described by the following $2\times2$ Hamiltonian,
$H(\textbf{k},m)=f_{0}(\textbf{k},m)+\sum_{i=1}^{3}f_{i}(\textbf{k},m)\tau_{i}$,
where $f_{0,1,2,3}$ are real functions and $\tau_{1,2,3}$ are Pauli matrices indicating
the two bands.
Here $m$ describes a tuning
parameter. In particular, we consider the following situation.
For $m<m_{c}$, the system is fully gapped. An isolated BTP
occurs at the critical point $(\textbf{k},m)=(\textbf{k}_{c},m_{c})$ where $f_{1,2,3}(\textbf{k}_{c},m_{c})=0$.
Since $f_{0}$ does not affect ABC, we can neglect $f_{0}$.
Then the next question
is what happens when $m>m_{c}$.
To examine the system's behavior near the critical point,
we derive the effective Hamiltonian through an expansion in powers of $\textbf{q}=\textbf{k}-\textbf{k}_{c}$ and $\Delta m=m-m_{c}$.
Up to the linear order of $\textbf{q}$ and $\Delta m$, $\textbf{f}=(f_{1},f_{2},f_{3})^{T}$
($T$ stands for transpose) can be written as
$\textbf{f}(\textbf{q},\Delta m)=\hat{M}\textbf{q}+\Delta m \textbf{N}$
where $\hat{M}_{ij}=\frac{\partial f_{i}}{\partial q_{j}}|_{\textbf{q}=\Delta m=0}$
and $N_{i}=\frac{\partial f_{i}}{\partial m}|_{\textbf{q}=\Delta m=0}$.
If the determinant of $\hat{M}$, i.e., $\text{Det}\hat{M}$, is nonzero,
the gap-closing condition $\textbf{f}=0$ leads to
$\textbf{q}=-\hat{M}^{-1}\textbf{N}\Delta m$,
which means that the gapless point moves as $\Delta m$ varies and persists even when $\Delta m<0$, contradicting
the initial assumption. Therefore $\text{Det}\hat{M}$=0 at the PT point.
In fact, the sign of $\text{Det}\hat{M}=\varepsilon_{ijk}M_{1i}M_{2j}M_{3k}$ is
the chirality (or chiral charge) of the BTP at $\Delta m=0$.
Since the chirality is a topological number, a BTP
with a nonzero chirality is stable against small perturbations.
However, when $\text{Det}\hat{M}$=0,
it is not topologically protected. Therefore when $\Delta m>0$,
the BTP can either be gapped out leading to another insulating phase or be split into
several Weyl points with zero net chirality generating a stable metallic phase.
When both of these possibilities are allowed, the insulating phase should be preferred
since the gapped phase has lower energy.

To understand the nature of the ground state for $\Delta m>0$, it is useful
to rotate the momentum coordinate using a basis which manifests the zero chirality
of the BTP at $\Delta m=0$.
Since $\text{Det}\hat{M}$=0, $\hat{M}$ has an eigenvector $\textbf{n}_{1}$ with zero eigenvalue
satisfying $\hat{M}\textbf{n}_{1}=0$. We introduce two additional normalized vectors $\textbf{n}_{2}$, $\textbf{n}_{3}$,
which can form an orthonormal basis $\{\textbf{n}_{1},\textbf{n}_{2},\textbf{n}_{3}\}$,
and construct a matrix $\hat{W}=(\textbf{n}_{1},\textbf{n}_{2},\textbf{n}_{3})$.
With the rotated coordinate $\textbf{p}=\hat{W}^{-1}\textbf{q}$,
$\textbf{f}(\textbf{p},\Delta m)=\textbf{u}_{2}p_{2}+\textbf{u}_{3}p_{3}+\Delta m \textbf{N}$,
where $\textbf{u}_{2,3}=\hat{M}\textbf{n}_{2,3}$.
Here terms linear in $p_{1}$ do not appear
in $\textbf{f}$ due to the fact that $\hat{M}\textbf{n}_{1}=0$.
Then the leading contribution of $p_{1}$ dependent term should start from quadratic order,
which leads to the minimal effective Hamiltonian $H(\textbf{p},\Delta m)=\sum_{i=1}^{3}f_{i}(\textbf{p},\Delta m)\tau_{i}$ in which
\begin{align}\label{eqn:H_minimal}
\textbf{f}(\textbf{p},\Delta m)=\textbf{u}_{2}p_{2}+\textbf{u}_{3}p_{3}+\textbf{u}_{4}p_{1}^{2}+\Delta m \textbf{N}.
\end{align}

\textit{Conditions to obtain an insulator.-}
Let us first derive the condition for IIT corresponding to
Fig.~\ref{fig:phasediagram} (a).
Since the system is gapped for any $\Delta m\neq 0$, the conduction (valence) band should
have a well-defined dispersion minimum (maximum) near $\textbf{p}=0$.
Considering the minimal $2\times 2$ Hamiltonian with $f_{1,2,3}(\textbf{p},\Delta m)$ in Eq.~(\ref{eqn:H_minimal}),
the condition to have an extremum for small $\Delta m\neq 0$ leads to
the following three equations $g_{i}=\partial E_{c}(\textbf{p},\Delta m)/\partial p_{i}=0$ $(i=1,2,3)$.
Here $E_{c}$ is the energy of the conduction band.
After solving the 3 coupled equations,
the location of the dispersion minimum is obtained as  $\textbf{p}^{\text{min}}=(0,A\Delta m,B\Delta m)$
where $A, B$ are some constants.
This implies that across the ABC,
the conduction (valence) band minimum (maximum) should move along the straight line satisfying
$p_{1}=0$
and $p_{2}=\frac{A}{B}p_{3}$ for both $\Delta m<0$ and $\Delta m>0$.
Such a condition can be satisfied generally when the system has high crystalline symmetry along the line.
Therefore the IIT is achievable when the extrema of the conduction and valence bands
of the gapped phases move along a straight line
across the ABC.

As a consequence of the IIT,
the energy dispersion develops a peculiar structure.
To understand the band shape near the dispersion minimum, we compute
the Hessian matrix $\hat{H}^{\text{min}}_{ij}=\frac{\partial^{2}E_{c}}{\partial p_{i}\partial p_{j}}$,
which has a block diagonal form with $\hat{H}^{\text{min}}_{12}=\hat{H}^{\text{min}}_{13}=0$
at $\textbf{p}=\textbf{p}^{\text{min}}$. The other nonzero components of $\hat{H}^{\text{min}}$ satisfies
\begin{displaymath}\label{eqn:Hessian1}
\hat{H}^{\text{min}}_{11}=c_{11} \Delta m,\quad
\text{Det}
\left( \begin{array}{cc}
H^{\text{min}}_{22} & H^{\text{min}}_{23} \\
H^{\text{min}}_{32} & H^{\text{min}}_{33}
\end{array} \right)
>0,
\end{displaymath}
where $c_{11}$ is a constant. Interestingly,
$\hat{H}^{\text{min}}_{11}$ changes the sign across the PT because
it is linearly proportional to $\Delta m$.
For $c_{11}<0$ ($c_{11}>0$), the conduction band
has a dispersion minimum in all three directions for $\Delta m <0$ ($\Delta m >0$)
while it has a saddle point with a negative curvature along the $p_{1}$ direction
for $\Delta m >0$ ($\Delta m <0$).
Therefore when there is a IIT,
one insulating phase should possess a saddle point
at the bottom (top) of the conduction (valence) band along the $p_{1}$ direction where the energy
dispersion is quadratic at QCP.

\begin{figure}[t]
\centering
\includegraphics[width=3in]{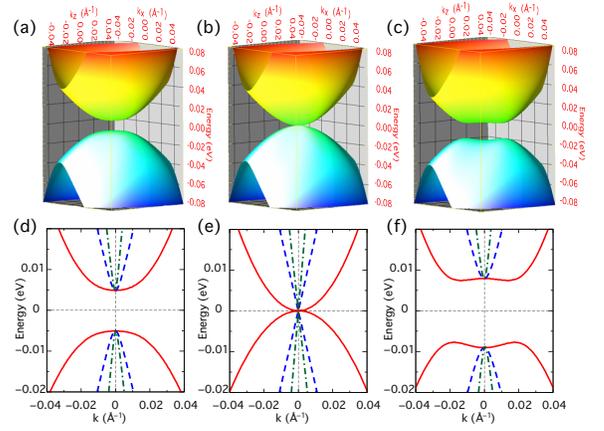}
\caption{(Color online)
Evolution of the band structure, obtained from first-principle calculations, across the topological PT
in BiTeI under pressure $P$. Energy dispersion of the conduction/valence bands near
one of the BTP in $(p_{1},p_{2})$ plane, which is normal to the high symmetry line
embracing QCPs, is shown
for (a) $P<P_{c}$, (b) $P=P_{c}$, (c) $P>P_{c}$, respectively.
Energy dispersions along the $p_{1}$ (red), $p_{2}$ (green), $p_{3}$ (blue) directions
are shown for (d) $P<P_{c}$, (e) $P=P_{c}$, and (f) $P>P_{c}$, respectively.
} \label{fig:dispersion}
\end{figure}

We can apply this theory to the IIT
of the pressured BiTeI.~\cite{BiTeI_theory2}.
In this system,
ABC occurs along the high symmetry line $A$-$H$ in $k_{z}=\pi$ plane (BZ of BiTeI is shown in Fig. 1 of ~\cite{BiTeI_theory2}).
Because of the $C_{3v}$ symmetry, the conduction (valence) band
with Rashba-type spin-splitting develops a dispersion minimum (maximum) along this line
for any pressure across the ABC, which satisfies the necessary condition
for the IIT.
In Fig.~\ref{fig:dispersion},
we plot the evolution of the band dispersion across the ABC
near one of BTPs
using the band structure obtained by first principle calculations.~\cite{DFT}
At the QCP, the band dispersion is quadratic along one direction and
linear along the other two directions.
Moreover, beyond the critical pressure, the band dispersion of
the insulating phase possesses a saddle point,
proving the occurrence of the IIT.

\textit{Conditions to obtain a semi-metal.}$-$
If the condition for gap reopening is not satisfied,
the BTP at $\Delta m=0$ can be split
into several BTPs.
Here we focus on the case of generating two BTPs with opposite chiral charges
for convenience.
Since there are 4 parameters ($p_{1,2,3}$ and $\Delta m$) while
only 3 conditions of $f_{1,2,3}=0$ are required to be satisfied
to achieve a gapless phase, there is a line of
gapless points in the ($\textbf{p}$, $\Delta m$) space in general.
Regarding $t\equiv\Delta m$ as a parameter, the trajectories of the two BTPs
form a curve $\textbf{p}^{*}(t)=(p_{1}^{*}(t),p_{2}^{*}(t),p_{3}^{*}(t))$ in 3D momentum space.
To determine the structure of the phase diagram, it is
crucial to understand the shape
of the curve in 3D space.

\begin{figure}[t]
\centering
\includegraphics[width=3 in]{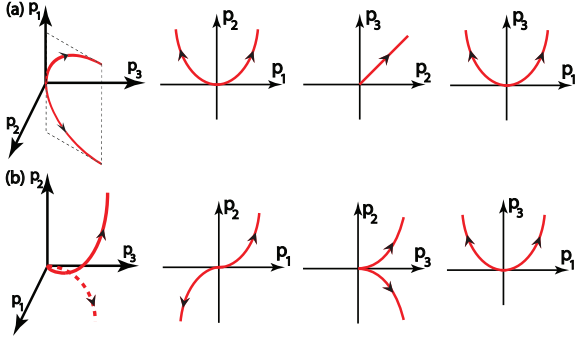}
\caption{(Color online)
The trajectory of BTPs in 3D space and
its 2D projections.
(a) The curve is lying on a 2D plane leading to Fig.~\ref{fig:phasediagram} (b).
(b) The curve is moving in 3D space leading to Fig.~\ref{fig:phasediagram} (c).
} \label{fig:curve}
\end{figure}

When the BTP, $\textbf{p}^{*}(t=0)$ is free of symmetry constraints,
the three components of $\textbf{p}^{*}(t=0)$ are linearly independent in general.
In this case, from Eq.~(\ref{eqn:H_minimal}), the location of
BTPs for small $t>0$ can be obtained as
$\textbf{p}^{*}(t)=(\pm a_{1}\sqrt{t},a_{2}t,a_{3}t)$ with $a_{1,2,3}$ constants,
which is initially proposed by Murakami and Kuga in Ref.~\onlinecite{Murakami}.
The shape of this trajectory in 3D momentum space and its 2D projections
are shown in Fig.~\ref{fig:curve} (a).
Since the curve is lying on a 2D plane,
the trajectory can form a closed loop, which can generate
another insulating state via a pair-annihilation of BTPs.
Therefore an ABC at a generic momentum point without symmetry constraints
can give rise to Fig.~\ref{fig:phasediagram} (b).~\cite{Murakami}

On the other hand, when $\textbf{p}^{*}(t=0)$
is under symmetry constraints,
the components of $\textbf{p}^{*}(t=0)$ cannot be linearly independent.
For example, in BiTeI,
$\textbf{p}^{*}(t=0)$ exists on a line where
the Hamiltonian is invariant under the combination of
time-reversal and mirror symmetries.
Although the IIT should occur in this system,
let us suppose that the splitting of the BTP is possible.
In this case,
it can be shown that the trajectory
follows $\textbf{p}^{*}(t)=(\pm\alpha\sqrt{t},\pm\beta t^{3/2},\gamma t)$ with constants $\alpha$, $\beta$, $\gamma$.
This is because the components $p_{1,2,3}$ of $\textbf{p}^{*}(t)$
satisfy $p_{2}\propto p_{1}p_{3}$ due to the symmetry constraint at $t=0$.
The detailed derivation
is provided in the Supplementary Material.
The shape of this trajectory is shown in Fig.~\ref{fig:curve} (b).
It is worth noting that the trajectory moves
in 3D space. It is vanishingly improbable that
two curves emanating from the origin and traveling in 3D space can collide again considering
the huge volume of the momentum space. Therefore if an ABC occurs
at a momentum under symmetry constraints, the trajectory of BTPs
can form an open curve leading to the phase diagram in Fig.~\ref{fig:phasediagram} (c).

\textit{Topological PT.}$-$
The IIT
can accompany the change of bulk topological properties.~\cite{TPT}
In 3D systems with TRS, band insulators can be classified
by $Z_{2}$ topological numbers $\nu_{0,1,2,3}$.~\cite{Fu-Kane-Mele,MooreBalents,Roy}
In the BZ, there are 3 pairs of parallel planes, in which
$\textbf{k}\cdot \textbf{a}_{i}=0$ or $\pi$. ($i=1,2,3$)
Here $\textbf{a}_{1,2,3}$ are primitive lattice vectors.
Since each plane has TRS,
a 2D $Z_{2}$ invariant $\alpha^{0}_{i}$ ($\alpha^{\pi}_{i}$) can be assigned to the plane satisfying
$\textbf{k}\cdot \textbf{a}_{i}=0$ ($\textbf{k}\cdot \textbf{a}_{i}=\pi$).
Since $\alpha^{0}_{1}+\alpha^{\pi}_{1}=\alpha^{0}_{2}+\alpha^{\pi}_{2}=\alpha^{0}_{3}+\alpha^{\pi}_{3}$,
only four 2D invariants are independent and determine the $Z_{2}$ invariants
of the 3D system in the following way,
$(\nu_{0},\nu_{1},\nu_{2},\nu_{3})=(\alpha^{0}_{1}+\alpha^{\pi}_{1},\alpha^{\pi}_{1},\alpha^{\pi}_{2},\alpha^{\pi}_{3})$.
The strong invariant $\nu_{0}$ distinguishes a TI ($\nu_{0}=1$)
and a band insulator ($\nu_{0}=0$).
Since $\nu_{0}=\alpha^{0}_{i}+\alpha^{\pi}_{i}$ for any $i=1,2,3$,
if one of 2D $Z_{2}$ invariants changes by 1 through ABC,
topological PT occurs.

In a 2D BZ with TRS, the $Z_{2}$ invariant $\alpha$
is given by the Chern number (modulo 2), which is the integral of
the Berry curvature over the half BZ (with additional contraction procedures).~\cite{MooreBalents}
Therefore if the ABC between the valence and conduction bands,
changing the Chern number of each band by $\pm1$ per a touching~\cite{Oshikawa},
occurs odd number of times in the half BZ, $\alpha$ changes by 1 leading to
3D topological PT.
Therefore when IIT happens,
if the high crystalline symmetry line embracing QCPs
is on a 2D plane with TRS
and the number of such lines in the half BZ is odd,
a topological PT occurs.
This condition is exactly satisfied in BiTeI where
three high symmetry lines embracing BTPs
are on the $k_{z}=\pi$ plane with TRS
leading to the topological PT.~\cite{BiTeI_theory2}

\textit{Thermodynamic properties at QCP.-}
The anisotropic dispersion of the BTP with zero chirality
leads to the following
minimal Hamiltonian at the QCP,
\begin{align}\label{eqn:Hamiltonian_QCP}
H_{\text{QCP}}(\textbf{p})=Ap_{1}^{2}\tau_{1}+\upsilon p_{2}\tau_{2}+\upsilon p_{3}\tau_{3}.
\end{align}
where $\upsilon$ is the velocity and $A$ is the inverse mass.
This gives rise to
the density of states $D(\varepsilon)\propto\varepsilon^{3/2}$, which is quite distinct from
that for a 3D Weyl semi-metal ($D(\varepsilon)\propto\varepsilon^{2}$) or a 3D normal metal with quadratic
dispersion ($D(\varepsilon)\propto\varepsilon^{1/2}$). The distinct power law of $D(\varepsilon)$
directly leads to new exponents in the temperature dependence of various thermodynamic quantities such
as the specific heat ($C_{V}$) and compressibility ($\kappa$) as summarized in Table~\ref{table:thermodynamic}.
\begin{table}
\begin{tabular}{@{}|c|c|c|c|c|c|}
\hline \hline
& $D(\varepsilon)$ & $C_{V}(T)$ & $\kappa(T)$ & $\chi_{D}(T)$ & $\sigma_{DC}(T)$\\
\hline \hline
$\text{Weyl semi-metal}$ & $\varepsilon^{2}$ & $T^{3}$ & $T^{2}$ & $\ln T$ & $T$
\\
\hline
$\text{At the QCP}$ & $\varepsilon^{3/2}$ & $T^{5/2}$ & $T^{3/2}$ & $T^{-1/2}$ & $T^{1/2}$
\\
\hline \hline
\end{tabular}
\caption{Temperature (or energy) dependence of
various physical quantities for a 3D Weyl semi-metal and at the QCP.
$D(\varepsilon)$, $C_{V}$, $\kappa$, $\chi_{D}$, $\sigma_{DC}$
are the density of states, specific heat, compressibility, diamagnetic susceptibility,
and DC conductivity, respectively.
$\sigma_{DC}(T)$ is obtained by using the T-linear scattering rate
due to Coulomb interaction between electrons.
}
\label{table:thermodynamic}
\end{table}
The diamagnetic susceptibility $\chi_{D}$ also shows
an unexpected singular behavior.
We have computed $\chi_{D}$ using the Fukuyama formula for the orbital susceptibility
$\chi_{D}=\frac{e^{2}}{c^{2}}\frac{T}{V}\sum_{n,\textbf{p}}\text{Tr}[G\gamma_{a}G\gamma_{b}G\gamma_{a}G\gamma_{b}]$.~\cite{Fukuyama}
Here $G$ is the Green's function,
$\gamma_{a}\equiv\frac{\partial H_{\text{QCP}}}{\partial p_{a}}$ and $a, b$ are two orthogonal directions
perpendicular to the applied magnetic field.
From Eq.~(\ref{eqn:Hamiltonian_QCP}), $\chi_{D}$ is given by
$\chi_{D}(\theta)=\cos^{2}\theta \chi_{1}+\sin^{2}\theta \chi_{2}$,
in which $\chi_{1}\approx C_{1}T^{-1/2}$ and $\chi_{2}\approx \chi_{2}^{0}+C_{2}T^{1/2}$
with $\chi_{2}^{0}$, $C_{1,2}$ constants.
Here $\theta$ is the angle between the external magnetic field and $p_{1}$ direction.
Therefore $\chi_{D}$ shows unusual singular temperature dependence in low temperature
given by $\chi_{D}\sim T^{-1/2}$ irrespective of magnetic field directions.

\textit{Anisotropic DC conductivity.-}
The anisotropic dispersion at QCP also induces anisotropic temperature
dependence of the DC conductivities.
Assuming momentum independence of the scattering rate $\frac{1}{\tau(\omega)}$,
a straightforward calculation of the conductivity tensor using Kubo formula
gives rise to the following expression
of the DC conductivities,
\begin{align}\label{eqn:DCconductivity}
\sigma_{11}(T)&=\frac{2e^{2}\sqrt{A}}{7\pi^{2}\upsilon^{2}}\int d\omega |\omega|^{5/2}\Big(-\frac{\partial f}{\partial\omega}\Big)\tau(\omega),
\nonumber\\
\sigma_{22,33}(T)&=\frac{9e^{2}}{20\pi^{2}\sqrt{A}}\int d\omega |\omega|^{3/2}\Big(-\frac{\partial f}{\partial\omega}\Big)\tau(\omega),
\end{align}

When the Coulomb interaction between electrons dominates the scattering,
we can take $\frac{1}{\tau}=\alpha^{2}T$ with $\alpha=\frac{e^{2}}{4\pi\varepsilon \upsilon}$,
considering that the low temperature transport is dominated by
the linear dispersion.
In this case, the DC conductivity satisfies
$\sigma_{11}(T)\propto T^{3/2}$ and
$\sigma_{22,33}(T)\propto T^{1/2}$.
On the other hand, when the scattering due to random potentials
dominates the transport, using Born approximation, the leading contribution to the scattering rate can be obtained by
$\frac{1}{\tau(w)}
\approx 2\pi\gamma_{0}D(w)$ with $\gamma_{0}=\frac{n_{i} V_{0}^{2}}{2}$.~\cite{disorder}
Here $V_{0}$ is
the impurity scattering potential, $n_{i}$ is the impurity density.
Then using Eq.~($\ref{eqn:DCconductivity}$), we obtain
$\sigma_{33}(T)=\frac{9e^{2}\upsilon^{2}}{20\pi\gamma_{0}}$,
$\sigma_{11}(T)=\frac{2e^{2}A}{7\pi\gamma_{0}}(2\ln2)T$, which also
shows the anisotropic $T$ dependence.~\cite{disorder_conductivity}
In fact, Eq.~($\ref{eqn:DCconductivity}$) implies that, as long as
the scattering rate is momentum independent, irrespective of the scattering mechanism
$\frac{\sigma_{11}(T)}{\sigma_{33}(T)}=C_{0}\frac{A}{\upsilon^{2}}T$ where
$C_{0}\approx 1.8$.

\textit{Stability of QCP.-}
Finally, let us discuss about the stability of the QCP
against disorder and Coulomb interaction.
The effective action of the QCP including both random disorder potential and $1/r$
Coulomb interaction can be written as
\begin{align}
S=&\int dt d^{3}x[\psi^{\dag}(i\partial_{t}+A\partial_{1}^{2}\tau_{1}+\sum_{j=2,3}i\upsilon\partial_{j}\tau_{j})\psi+V_{i}\psi^{\dag}M_{i}\psi]
\nonumber\\
&+\int dt d^{3}xd^{3}x'(\psi^{\dag}\psi)_{x,t}\frac{g^{2}}{2|\textbf{x}-\textbf{x}'|}(\psi^{\dag}\psi)_{x',t}.
\end{align}
where $V_{i}(\textbf{x})$ is a random potential coupled to fermion field $\psi(\textbf{x})$
via a matrix $M_{i}$. $g^{2}=e^{2}/\varepsilon$ where $e$ and $\varepsilon$ are the electric charge and dielectric constant,
respectively. We take a random disorder potential with Gaussian invariance whose
impurity average satisfies $\langle V_{i}(\textbf{x})V_{j}(\textbf{x}')\rangle=\Delta_{ij}\delta^{(3)}(\textbf{x}-\textbf{x}')$.
The key characteristics of the Gaussian fixed point in Eq.~(\ref{eqn:Hamiltonian_QCP})
is the invariance of the Hamiltonian under the anisotropic scaling of spatial
coordinates, i.e., $\tilde{x}_{1}= x_{1}/b^{1/2}$, $\tilde{x}_{2,3}= x_{2,3}/b$
accompanied by $\tilde{t}=t/b$ where the tilde indicates the new scaled coordinates.
Under this scale transformation, $\Delta_{ij}$ transforms as $\tilde{\Delta}_{ij}=b^{-1/2}\Delta_{ij}$
showing the irrelevance of the disorder.
Similarly, we can show that $\tilde{g}^{2}=g^{2}$, i.e.,
Coulomb interaction is marginal, which, however,
eventually becomes irrelevant according to
the one-loop perturbative renormalization group calculation.~\cite{RG}
Therefore the unusual power laws in various thermodynamic and transport properties, which are predicted
based on the free particle Hamiltonian in Eq.~(\ref{eqn:Hamiltonian_QCP})
should persist even under the influence of the disorder and Coulomb interaction.

We greatly appreciate the stimulating discussions with Ammon Aharony, Ora Entin-Wohlman and Michael Hermele.
This work is supported by the Japan Society for the Promotion
of Science (JSPS) through the ``Funding Program for World-Leading Innovative
R$\&$D on Science and Technology (FIRST Program)",
and by Grant-in-Aids for Scientific Research (No. 24224009) 
from the Ministry of Education, Culture, Sports, Science and Technology (MEXT) of Japan.



\section{\label{sec:supplement} Supplementary Material}

\subsection{\label{sec:curve} A curve in 3D and its curvature and torsion}

In this section, we describe the relation between
the shape of a curve in 3D space and its curvature and torsion.
Here we basically follow the contents in Ref.~\onlinecite{Spivak}.
A convenient way to describe a curve $C=C(t)=(x(t),y(t),z(t))$
is to use an orthogonal coordinate system $(\textbf{t},\textbf{n},\textbf{b})$
where $\textbf{t}$, $\textbf{n}$, $\textbf{b}$ are the tangential, normal, and binormal
vectors, respectively.
To define $\textbf{t}$, $\textbf{n}$, $\textbf{b}$ we first consider the arclength,
which is defined as
\begin{align}
s(t)=\int_{0}^{t}dt'\Big|\frac{dC(t')}{dt'}\Big|.
\end{align}
For the following discussion, we reparametrize the curve $C$ using the arclength $s$, which
makes $C$ to be a function of $s$, i.e.,$C=C(s)$.
Then $\textbf{t}$, $\textbf{n}$, $\textbf{b}$ are given by
\begin{align}\label{eqn:tnb1}
\textbf{t}\equiv \frac{dC}{ds},\quad
\textbf{n}\equiv \frac{d^{2}C}{ds^{2}} / \Big|\frac{d^{2}C}{ds^{2}}\Big|,\quad
\textbf{b}\equiv \textbf{t}\times \textbf{n},
\end{align}
and the curvature $\kappa_{C}$ and torsion $\tau_{C}$ are defined as
\begin{align}\label{eqn:tnb2}
\kappa_{C}(s)\equiv\Big|\frac{d\textbf{t}}{ds}\Big|,\quad \frac{d\textbf{b}}{ds}\equiv -\tau_{C}(s)\textbf{n}.
\end{align}
Therefore $\kappa_{C}$ measures the rate at which the tangential vector changes
and $\tau_{C}$ measures the rate at which the curve $C$ deviates from
being a planar curve lying on the $(\textbf{t},\textbf{n})$ space.
In the case of the torsion $\tau_{C}$, it has a following equivalent expression,
\begin{align}\label{eqn:torsion}
\tau_{C}=\frac{1}{\kappa_{C}^{2}}\Big\langle \frac{dC}{ds}\times\frac{d^{2}C}{ds^{2}},\frac{d^{3}C}{ds^{3}} \Big\rangle,
\end{align}
where $\langle A,B\rangle$ indicates the inner product of two vectors $A$ and $B$.

\begin{figure}[t]
\centering
\includegraphics[width=3 in]{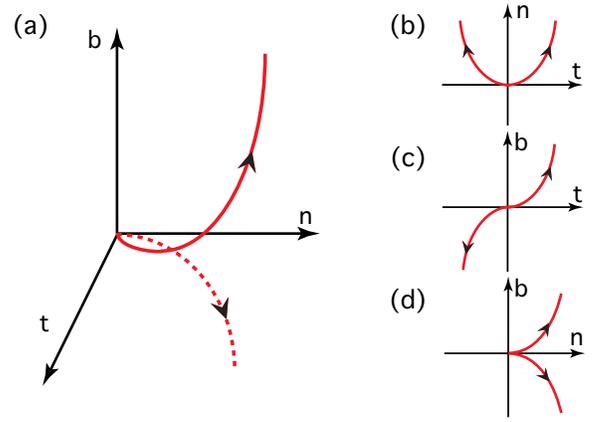}
\caption{(Color online)
(a)  Typical shape of a curve with nonzero curvature and torsion in 3D space.
2D Projections of the curve onto (b) (\textbf{t},\textbf{n}) plane,
(c) (\textbf{t},\textbf{b}) plane,
(d) (\textbf{n},\textbf{b}) plane.
Here two arrows describe the flowing directions of two band touching points
which are generated from the QCP at the origin.
} \label{fig:trajectory3D}
\end{figure}

The relation between the shape of a curve and its curvature and torsion can be understood
by considering two dimensional projections of the curve on the planes
in which two vectors among $\{\textbf{t}, \textbf{n}, \textbf{b}\}$ are adopted as a basis.
To represent an arbitrary curve $C(s)$ we can choose a coordinate
which satisfies $C(0)=0$, $\frac{dC}{ds}\Big|_{s=0}=(1,0,0)$, and $\frac{d^{2}C}{ds^{2}}\Big|_{s=0}=(0,\kappa_{C},0)$.
Then from Eq.~(\ref{eqn:torsion}),
we obtain $\frac{d^{3}C}{ds^{3}}\Big|_{s=0}=(a_{1},a_{2},\kappa_{C}\tau_{C})$ where $a_{1,2}$ are arbitrary numbers.

Now we consider the Taylor expansion of $C(s)=(x(s),y(s),z(s))$ at $s=0$,
\begin{align}
C(s)=C(0)+s\frac{dC}{ds}\Big|_{0}+\frac{s^{2}}{2}\frac{d^{2}C}{ds^{2}}\Big|_{0}
+\frac{s^{3}}{6}\frac{d^{3}C}{ds^{3}}\Big|_{0}+O(s^{4}),
\end{align}
which gives rise to
\begin{align}
x(s)&=s+O(s^{3}),
\nonumber\\
y(s)&=\frac{\kappa_{C}}{2}s^{2}+O(s^{3}),
\nonumber\\
z(s)&=\frac{\kappa_{C}\tau_{C}}{6}s^{3}+O(s^{4}).
\end{align}
From the leading order terms, the projections of the curve $C$ are described by
$y=\frac{\kappa_{C}}{2}x^{2}$, $z=\frac{\kappa_{C}\tau_{C}}{6}x^{3}$, and
$z^{2}=\frac{2}{9}\frac{\tau^{2}_{C}}{\kappa_{C}}y^{3}$,
which are lying on $(\textbf{t},\textbf{n})$, $(\textbf{t},\textbf{b})$, and $(\textbf{n},\textbf{b})$ planes, respectively.
A typical example of a curve in 3D and its 2D projections are shown in Fig.~\ref{fig:trajectory3D}.
It is to be noted that the projections of the curve on
$(\textbf{t},\textbf{b})$ and $(\textbf{n},\textbf{b})$ planes,
shown in Fig.~\ref{fig:trajectory3D} (c) and (d), respectively,
cannot form a closed curve under smooth variations of the curve.
Since the shapes of these 2D projections remain the same as long as
the torsion is nonzero at the origin,
we obtain the zero torsion condition $\tau_{C}=0$ to form a closed curve.
On the other band, when $\tau_{C}=0$,
the curve moves on the 2D space spanned by $(\textbf{t},\textbf{n})$.
The trajectory of the curve on $(\textbf{t},\textbf{n})$ plane shown in Fig.~\ref{fig:trajectory3D} (b)
can make a closed loop as long as the curvature $\kappa_{C}$ is finite.

\subsection{\label{sec:effect_H} Effective Hamiltonian for topological phase transition in BiTeI}

Let us first consider a general Hamiltonian defined in momentum space $H(\textbf{k})=H(k_{x},k_{y},k_{z})$.
Then the combined symmetry operation $TM$, which is the combination of the time reversal ($T$) and mirror ($M:y\rightarrow -y$),
imposes the following constraint to the $H(\textbf{k})$,
\begin{align}
&(TM)H(k_{x},k_{y},k_{z})(TM)^{-1}
\nonumber\\
&=H^{*}(k_{x},k_{y},k_{z})=H(-k_{x},k_{y},-k_{z}).
\end{align}
Assuming that the band touching point exists at $\textbf{k}_{c}=(0,k_{y,c},\pi)$,
we derive the low energy Hamiltonian considering small momentum deviation from the band touching point.
Due to the $TM$ symmetry, the effective $2\times 2$ Hamiltonian
$H_{2\times 2}(\textbf{q})$ with $\textbf{q}=\textbf{k}-\textbf{k}_{c}$, satisfies
the following constraint,
\begin{align}
&(TM)H_{2\times 2}(q_{x},q_{y},q_{z})(TM)^{-1}
=H_{2\times 2}^{*}(q_{x},q_{y},q_{z})
\nonumber\\
&=H_{2\times 2}^{*}(k_{x},k_{y}-k_{y,c},k_{z}-\pi)
\nonumber\\
&=H_{2\times 2}(-k_{x},k_{y}-k_{y,c},-k_{z}-\pi)
\nonumber\\
&=H_{2\times 2}(-k_{x},k_{y}-k_{y,c},-k_{z}+\pi)
\nonumber\\
&=H_{2\times 2}(-q_{x},q_{y},-q_{z}).
\end{align}
Therefore
\begin{align}\label{eqn:constraint}
H_{2\times 2}^{*}(q_{x},q_{y},q_{z})=H_{2\times 2}(-q_{x},q_{y},-q_{z})
\end{align}
For the generic $2\times 2$ Hamiltonian $H_{2\times 2}(\textbf{q})$ given by,
\begin{align}
H_{2\times 2}(\textbf{q})=f_{1}(\textbf{q})\tau_{1}+f_{2}(\textbf{q})\tau_{2}+f_{3}(\textbf{q})\tau_{3},
\end{align}
the constraint in Eq.~(\ref{eqn:constraint}) leads to the following constraints to $f_{1,2,3}(\textbf{q})$,
\begin{align}\label{eqn:fconstraint}
f_{1,3}(-q_{x},q_{y},-q_{z})&=f_{x,z}(q_{x},q_{y},q_{z})
\nonumber\\
f_{2}(-q_{x},q_{y},-q_{z})&=-f_{y}(q_{x},q_{y},q_{z})
\end{align}
which means that $f_{1,3}$ ($f_{2}$) are even (odd) under the simultaneous sign change of $q_{x}$ and $q_{z}$.
Now we expand $f_{1,2,3}$ near the band touching point at $\textbf{q}=0$ and $P=P_{c}$
in the powers of $q_{x,y,z}$ and $\Delta P=P-P_{c}$.
Due to the symmetry constraint in Eq.~(\ref{eqn:fconstraint}), up to linear order in
$q_{x,y,z}$ and $\Delta P=P-P_{c}$, $f_{1,2,3}$ are given by
\begin{align}
f_{1}&=N_{1}\Delta P + M_{12} q_{y}
\nonumber\\
f_{2}&=M_{21}q_{x} + M_{23} q_{z}
\nonumber\\
f_{3}&=N_{3}\Delta P + M_{32} q_{y}
\end{align}
where $N_{1,3}$ and $M_{12,21,23,32}$ are constants.
The Hamiltonian at the critical point $\Delta P=0$ can be written as
\begin{align}\label{eqn:Hcritical}
H_{2\times 2}(\textbf{q})=M_{12} q_{y}\tau_{1}+(M_{21}q_{x} + M_{23} q_{z})\tau_{2}+M_{32} q_{y}\tau_{3},
\end{align}
Since $\text{Det}M=0$, $M$ always has an eigenvector $\xi_{1}$ with zero eigenvalue. Explicitly,
$\xi_{1}^{t}=\frac{1}{\sqrt{M_{21}^{2} + M_{23}^{2}}}(-M_{23},0,M_{21})$
where the superscript $t$ means the transpose of a vector.

Let us introduce $\xi_{2}^{t}=\frac{1}{\sqrt{M_{21}^{2} + M_{23}^{2}}}(M_{21},0,M_{23})$ and $\xi_{3}^{t}=(0,1,0)$.
Then $\xi_{1,2,3}$ constitute an orthogonal basis. Using this, we consider following linear transformation,
\begin{align}\label{eqn:lineartransform}
\left( \begin{array}{c}
q_{x} \\
q_{y} \\
q_{z}
\end{array} \right)
&\equiv
\left( \begin{array}{ccc}
\xi_{1}, & \xi_{2}, & \xi_{3} \\
\end{array} \right)
\left( \begin{array}{c}
p_{1} \\
p_{2} \\
p_{3}
\end{array} \right)
\nonumber\\
&=
\left( \begin{array}{c}
\frac{-M_{23}}{\sqrt{M_{21}^{2} + M_{23}^{2}}} p_{1} +\frac{M_{21}}{\sqrt{M_{21}^{2} + M_{23}^{2}}} p_{2}\\
p_{3} \\
\frac{M_{21}}{\sqrt{M_{21}^{2} + M_{23}^{2}}} p_{1} +\frac{M_{23}}{\sqrt{M_{21}^{2} + M_{23}^{2}}} p_{2}
\end{array} \right)
\end{align}
Applying the above linear transformation, the Hamiltonian is given by
$H_{2\times 2}(\textbf{p})=\sum_{i=1,2,3}f_{i}(\textbf{p})\tau_{i}$ in which
\begin{align}\label{eqn:H_normalcoord}
f_{1}&=N_{1}\Delta P+M_{12} p_{3},
\nonumber\\
f_{2}&=\sqrt{M_{21}^{2} + M_{23}^{2}}p_{2},
\nonumber\\
f_{3}&=N_{3}\Delta P+ M_{32} p_{3}.
\end{align}
It is to be noticed that $p_{1}$ does not appear in the Hamiltonian due to the fact the $\xi_{1}$ is the
eigenvector of $M$ with zero eigenvalue.

To fully account for the phase transition, the terms quadratic in $p_{i}$
are necessary. In terms of the rotated momentum $\textbf{p}$,
the symmetry constraint in Eq.~(\ref{eqn:fconstraint}) can be written as
\begin{align}\label{eqn:constraint_p}
H_{2\times 2}^{*}(p_{1},p_{2},p_{3})=H_{2\times 2}(-p_{1},-p_{2},p_{3}).
\end{align}
Collecting terms satisfying the constraint above up to the quadratic order in $p_{i}$,
$f_{1,2,3}$ can be written as
\begin{align}\label{eqn:fquadratic_p2}
f_{1}&=N_{1}\Delta P + M_{12} p_{3} + a_{1}p_{1}^{2}+a_{2}p_{2}^{2}+a_{3}p_{3}^{2}+a_{4}p_{1}p_{2},
\nonumber\\
f_{2}&=\sqrt{M_{21}^{2}+ M_{23}^{2}} p_{2}+b_{5}p_{2}p_{3}+b_{6}p_{3}p_{1},
\nonumber\\
f_{3}&=N_{3}\Delta P + M_{32} p_{3} + c_{1}p_{1}^{2}+c_{2}p_{2}^{2}+c_{3}p_{3}^{2}+c_{4}p_{1}p_{2},
\end{align}
where $a_{i}$, $b_{i}$, $c_{i}$ ($i=1,2,...,6$) are constants.

\subsection{\label{sec:TPT} Topological phase transition in BiTeI }
\begin{figure}[t]
\centering
\includegraphics[width=4.5 cm]{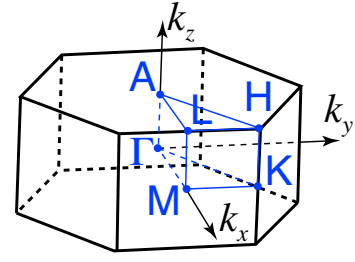}
\caption{(Color online)
The hexagonal Brillouin zone of BiTeI.
} \label{fig:BZ}
\end{figure}
Recently, it is shown that when external pressure is applied to BiTeI,
a direct phase transition from a normal insulator to a TI occurs
mediated by accidental band touching points at the critical pressure $P=P_{c}$.
The nature of the topological phase transition in BiTeI can be understood based
on the general considerations developed in previous discussions.
The key ingredient to understand the phase transition in this material is the fact
that in the BZ shown in Fig.~\ref{fig:BZ}, the band touching occurs along the $A-H$ direction in $k_{z}=\pi$ plane,
along which the Hamiltonian $H(\textbf{k})$ is invariant under the
combination of the time reversal ($T$) and mirror ($M$) symmetries.
Picking one of the $A-H$ direction along the $k_{y}$ axis,
the combined symmetry operation $TM$
imposes the following constraint to the Hamiltonian,
$H^{*}(k_{x},k_{y},k_{z})=H(-k_{x},k_{y},-k_{z})$
because the mirror $M$ changes $k_{y}$ to $-k_{y}$.
This symmetry constraint restricts the structure of the low energy Hamiltonian near
the gap-closing points. Explicitly, the effective Hamiltonian near one
of the band touching points can be written as $H(\textbf{p},\Delta P)=\sum_{i=1}^{3}f_{i}(\textbf{p},\Delta P)\tau_{i}$
in which
\begin{align}\label{eqn:fquadratic_p}
f_{1}&=N_{1}\Delta P + M_{12} p_{3} + a_{1}p_{1}^{2}+a_{2}p_{2}^{2}+a_{3}p_{3}^{2}+a_{4}p_{1}p_{2},
\nonumber\\
f_{2}&=\sqrt{M_{21}^{2}+ M_{23}^{2}} p_{2}+b_{5}p_{2}p_{3}+b_{6}p_{3}p_{1},
\nonumber\\
f_{3}&=N_{3}\Delta P + M_{32} p_{3} + c_{1}p_{1}^{2}+c_{2}p_{2}^{2}+c_{3}p_{3}^{2}+c_{4}p_{1}p_{2},
\end{align}
where $a_{i}$, $b_{i}$, $c_{i}$ ($i=1,2,...,6$) are constants and $\Delta P=P-P_{c}$.
$\textbf{p}$ is the rotated momentum coordinates adapted to manifest
the zero chirality of the band touching point. Therefore the terms linear in $p_{1}$
do not appear in $f_{1,2,3}$.
The detailed procedures to derive the above Hamiltonian is shown in the Supplementary Material.

Let us first check the conditions to obtain a semi-metallic phase
by finding the solution of $f_{1,2,3}=0$.
When $|\textbf{p}|\ll 1$, $f_{2}=0$ leads to,
$p_{2}\propto p_{3}p_{1}$.
\begin{figure}[t]
\centering
\includegraphics[width=8.5 cm]{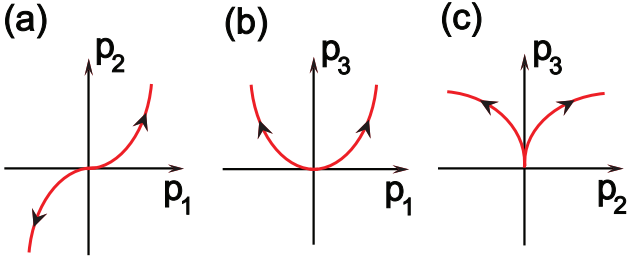}
\caption{(Color online)
The 2D projections of the curve satisfying the gap-closing condition
for the trajectory in Eq.~(\ref{eqn:trajectory_BiTeI})
for $\alpha>0$, $\beta>0$, $\gamma>0$.
} \label{fig:trajectory_BiTeI}
\end{figure}
Inserting this result to the conditions $f_{1,2}=0$,
the general solution of the gapless point has the following structure,
$p^{*}_{1}=\pm\alpha\sqrt{\Delta P}$, $p^{*}_{2}=\pm\beta(\Delta P)^{3/2}$, $p^{*}_{3}=\gamma \Delta P$
with $\alpha$, $\beta$, $\gamma$ constants.
Here we assume that the gapless point exists only for $\Delta P >0$.
Therefore the trajectory of the gapless points is given by
\begin{align}\label{eqn:trajectory_BiTeI}
p^{*}=(\pm\alpha\sqrt{\Delta P},\pm\beta(\Delta P)^{3/2},\gamma \Delta P)
\end{align}
whose curvature and torsion at $\Delta P=0$ is given by
$\kappa_{C}=\frac{2\gamma}{\alpha^{2}}$ and $\tau_{C}=\frac{-3\beta}{\alpha\gamma}$, respectively.
Since the curve has a finite torsion, we expect the trajectory of the gapless points
would not form a closed loop under the smooth variation of the system, which is supported by the corresponding
2D projections of the curve shown in Fig.~\ref{fig:trajectory_BiTeI}.
Therefore once a semi-metallic phase occurs by splitting the band touching point at $\Delta P=0$,
the gapless point should persist for all $\Delta P$, which is not consistent with
the predictions of the first principle calculation.

The only way to describe the transition between two insulators
is the occurrence of the direct transition between them via band touching at a single critical point.
From the condition that the conduction band has a minimum,
we have three equations of $\frac{\partial E_{c}(\textbf{p})}{\partial p_{i}}=0$ ($i=1,2,3$).
After careful examination of the three coupled equations,
we have found that there is a unique solution given by
\begin{align}
\textbf{p}^{\text{min}}=(0,0,-\frac{N_{1}M_{12}+N_{3}M_{32}}{M_{12}^{2}+M_{32}^{2}}\Delta P).
\end{align}
Therefore the direct transition between two gapped phases requires that
the extrema of the valence and conduction bands move along
a particular direction, the $p_{3}$ direction throughout the phase transition.
Interestingly, the $p_{3}$ direction corresponds to the $k_{y}$ direction
of the original coordinate, which is nothing but one of the $A-H$ direction in the BZ.
Because of the $C_{3v}$ point group symmetry, the conduction (valence) band
with the Rashba-type spin-orbit coupling develops a dispersion minimum (maximum) along the $A-H$ line
for any $\Delta P\neq0$. Moreover, since the $A-H$ line is on the $k_{z}=\pi$ plane satisfying
the time-reversal invariance, the band touching can induce the topological phase transition.
Explicitly, in the $k_{z}=\pi$ plane, the effective Hamiltonian
near a gap-closing point can be written as a two dimensional massive Dirac Hamiltonian,
in which $\Delta P$ plays the role of the mass term. Therefore a band touching point
can change the $Z_{2}$ invariant on the $k_{z}=\pi$ plane by 1 through the sign change of $\Delta P$.
Since there are 3 pairs of band touching points on the same plane,
the strong index $\nu_{0}$ changes by 1 reflecting the emergence
of the TI through the band touching.

The Hessian matrix $\hat{H}^{\text{min}}_{ij}=\frac{\partial^{2}E_{c}}{\partial p_{i}\partial p_{j}}$ has
a block diagonal form with $\hat{H}^{\text{min}}_{13}=\hat{H}^{\text{min}}_{23}=0$
at $\textbf{p}=\textbf{p}^{\text{min}}$. The other nonzero components of $\hat{H}^{\text{min}}$ satisfies
\begin{displaymath}\label{eqn:Hessian1}
\hat{H}^{\text{min}}_{33}>0,\quad
\text{Det}
\left( \begin{array}{cc}
H^{\text{min}}_{11} & H^{\text{min}}_{12} \\
H^{\text{min}}_{21} & H^{\text{min}}_{22}
\end{array} \right)
=C' \Delta P,
\end{displaymath}
where $C'$ is a constant.
Therefore the conduction band always has a minimum along the $p_{3}$ direction.
On the other hand, the determinant of the Hessian matrix for the $p_{1}$, $p_{2}$ directions,
normal to the $A-H$ line, shows sign
change across the phase transition, which predicts the appearance
of a saddle point in $(p_{1},p_{2})$ plane for one of the gapped phase.
Since the normal insulator for $\Delta P<0$ possesses the conventional dispersion
minimum or maximum, we can set $C'<0$.
Therefore the TI should possess saddle points in energy dispersion.
The change of the band dispersion across the topological phase transition
is explicitly shown
in Fig. 3 of the main text from the band structure obtained by first principle calculation.

\subsection{\label{sec:DC conductivity} DC conductivity at the quantum critical point}

Regarding the conductivity at the quantum critical point, there
are two important characteristics of BiTeI as compared to the isotropic 3D Dirac semi-metal system.
One is the enhanced density of states ($D(\epsilon)\propto\epsilon^{3/2}$),
which can be contrasted with the 3D Dirac system ($D(\epsilon)\propto\epsilon^{2}$),
and the other is the strong spatial anisotropy.

\textit{Influence of enhanced density of states.-}
Let us first consider the influence of the enhanced density of states using the semi-classical
Boltzmann theory. In the case of isotropic systems with constant scattering rates,
the longitudinal DC conductivity is given by
\begin{align}
\sigma=\frac{e^{2}\upsilon_{F}^{2}}{3}\int^{\infty}_{-\infty}
d\epsilon D(\epsilon)\Big[-\frac{\partial n_{F}(\epsilon)}{\partial \epsilon }\Big]\tau.
\end{align}
After plugging the density of states at the quantum critical point given by
\begin{align}
D(\epsilon)=\frac{1}{2\pi^{2}\upsilon^{2}\sqrt{A}}\epsilon^{3/2},
\end{align}
the conductivity can be obtained as
\begin{align}
\sigma&=\frac{e^{2}\upsilon_{F}^{2}}{3}\frac{1}{2\pi^{2}\upsilon^{2}\sqrt{A}}[\frac{3}{4}(2-\sqrt{2})\sqrt{\pi}\zeta(\frac{3}{2})](k_{B}T)^{3/2}\tau,
\nonumber\\
&\propto T^{3/2}\tau.
\end{align}
When the electron scattering is dominated by Coulomb interaction between electrons,
the scattering rate $\frac{1}{\tau}$
can be estimated in the following way,
\begin{align}
\frac{1}{\tau}\equiv -2 \text{Im}\Sigma^{\text{ret}}(k,\omega)\approx \alpha^{2}T.
\end{align}
The structure of $\tau$ from the Coulomb scattering can be understood in the following way.
At first, according to the Fermi-Golden rule, the scattering rate should be proportional to the
square of the scattering amplitude $\alpha$. In addition, since there is no energy scale other
than the temperature at the critical point, which immediately gives rise to the above form of $\tau$.
Then the temperature dependence of the DC conductivity is given by
\begin{align}
\sigma \propto T^{3/2}\tau = \frac{T^{1/2}}{\alpha^{2}},\qquad \rho \propto \frac{1}{T^{1/2}},
\end{align}
where $\rho$ is the longitudinal resistivity.
This can be contrasted with the corresponding quantities of the 3D Dirac semi-metal phase.~\cite{Hosur}
\begin{align}
\sigma \propto T^{2}\tau = \frac{T}{\alpha^{2}},\quad \rho \propto \frac{1}{T}.
\end{align}

\textit{ Anisotropic DC conductivity-.}
Now let us calculate the conductivity rigorously using Kubo formula considering
the anisotropic dispersion at the quantum critical point.
The effective Hamiltonian at the quantum critical point is given by
\begin{align}\label{eqn:Hamiltonian}
H(\textbf{k})&=A k_{1}^{2}\tau_{1}+\upsilon (k_{2}\tau_{2}+k_{3}\tau_{3}),
\end{align}
where $\tau_{1,2,3}$ are Pauli matrices representing the valence and conduction bands
that touch at the critical point. $\upsilon$ is the velocity and $A$ is the inverse mass
along the $k_{1}$ direction.

Kubo formula for frequency dependent conductivity is given by
\begin{align}
\sigma_{\mu\nu}(\omega,T)=-\frac{\text{Im} \Pi^{\text{ret}}_{\mu\nu}(\omega,T)}{\omega},
\end{align}
where
\begin{align}
\Pi_{\mu\nu}(i\nu_{n})=\frac{1}{\beta}\sum_{i\omega_{n}}\int\frac{d^{3}k}{(2\pi)^{3}}
\text{Tr}\Big[G_{\textbf{k},\omega_{n}+\nu_{n}}j_{\mu}(\textbf{k})G_{\textbf{k},\omega_{n}}j_{\nu}(\textbf{k}) \Big].
\end{align}
Here $G$ is the Matsubara Green's function and $j_{\mu}$ is the current operator
along the $\mu$ direction.
After some calculation, we can obtain the following expressions for current-current correlations,
\begin{align}
&\text{Im} \Pi^{\text{ret}}_{11}(\nu,T)=2e^{2}A^{2}\int\frac{d^{3}k}{(2\pi)^{3}}k_{1}^{2}\int\frac{d\varepsilon}{\pi}
[n_{F}(\varepsilon+\nu)-n_{F}(\varepsilon)]
\nonumber\\
&\times\sum_{\lambda,\lambda'}\text{Im}\mathcal{G}_{\lambda}^{\text{ret}}(k,\varepsilon+\nu)
\text{Im}\mathcal{G}_{\lambda'}^{\text{ret}}(k,\varepsilon)\Big\{1-\lambda\lambda'\frac{(\upsilon^{2}k^{2}_{\perp}-A^{2}k_{1}^{4})}{E^{2}}\Big\},
\end{align}
and
\begin{align}
&\text{Im} \Pi^{\text{ret}}_{22,33}(\nu,T)=\frac{e^{2}\upsilon^{2}}{2}\int\frac{d^{3}k}{(2\pi)^{3}}\int\frac{d\varepsilon}{\pi}
[n_{F}(\varepsilon+\nu)-n_{F}(\varepsilon)]
\nonumber\\
&\qquad\times\sum_{\lambda,\lambda'}\text{Im}\mathcal{G}_{\lambda}^{\text{ret}}(k,\varepsilon+\nu)
\text{Im}\mathcal{G}_{\lambda'}^{\text{ret}}(k,\varepsilon)\Big\{1-\lambda\lambda'\frac{A^{2}k_{1}^{4}}{E^{2}}\Big\},
\end{align}
where $\lambda=\pm$ indicates the positive/negative energy states and $E(\textbf{k})=\sqrt{\upsilon^{2}k_{\perp}^{2}+A^{2}k_{1}^{4}}$ with $k_{\perp}^{2}=k_{2}^{2}+k_{3}^{2}$.
Also the imaginary part of the retarded Green's function is given by
\begin{align}
\text{Im}\mathcal{G}_{\lambda}^{\text{ret}}(k,\varepsilon)=\frac{-\Gamma}{(\varepsilon-\lambda E(\textbf{k}))^{2}+\Gamma^{2}}.
\end{align}
After some complicate computations, we obtain the following expression of the DC conductivity due
to Coulomb interaction between electrons.
\begin{align}
\sigma_{11}(T)&=\frac{e^{2}\sqrt{A}}{7\pi^{2}\upsilon^{2}}\frac{1}{\Gamma}c_{11}(k_{B}T)^{5/2}\propto T^{3/2},
\nonumber\\
\sigma_{22,33}(T)&=\frac{9e^{2}}{40\pi^{2}\sqrt{A}}\frac{1}{\Gamma}c_{22,33}(k_{B}T)^{3/2}\propto T^{1/2},
\end{align}
where $c_{11,22,33}$ are constants.
It is to be noted that the in-plane conductivity ($\sigma_{22,33}$) follows the power law
expected for the density of states at the quantum critical point.
However, $\sigma_{11}(T)$ shows a completely different power law.

\textit{ DC conductivity due to disorder.-}
Let us study the conductivity at the quantum critical point due to charge-neutral point scatterers.
For this purpose, we compute the imaginary part of the electron self-energy due to
disorder using Born approximation.
The result is like the following.
\begin{align}
&\frac{1}{\tau_{\lambda}(\textbf{k},w)}\equiv -2\text{Im}\Sigma_{\lambda}^{\text{ret}}(\textbf{k},w)
\nonumber\\
=&\pi n_{i} V_{0}^{2}\Big(D(w)+\frac{Ak_{1}^{2}}{\upsilon^{2}k_{\perp}^{2}+A^{2}k_{1}^{4}}\frac{\lambda}{6\pi^{2}\upsilon^{2}A^{1/2}|w|^{3/2}\text{sgn}(w)}\Big)
\nonumber\\
=&\pi n_{i} V_{0}^{2}\frac{|\omega|^{3/2}}{2\pi^{2}\upsilon^{2}A^{1/2}}\Big(1+\frac{\lambda}{3}\text{sgn}(w)\frac{Ak_{1}^{2}}{\upsilon^{2}k_{\perp}^{2}+A^{2}k_{1}^{4}}\Big)
\end{align}
where $V_{0}$ is
the impurity scattering potential. $D(w)$ is the density of states.
In contrast to the isotropic systems, the scattering rate shows a momentum dependence.
However, we have to take into account of the fact that the linear dispersion dominates the low energy properties of the system
over the quadratic part, which is also supported by the one-loop
renormalization group calculation.~\cite{RG}
Therefore we can safely neglect the momentum dependence of the electron self-energy.
Then the scattering rate is just proportional to the density of states, which can be written as
\begin{align}
\frac{1}{\tau_{\lambda}(\textbf{k},w)}=\frac{1}{\tau(w)}
=&2\pi\gamma_{0}D(w),\quad \gamma_{0}=\frac{n_{i} V_{0}^{2}}{2},
\end{align}
which leads to
\begin{align}
\sigma_{11}(T)=\frac{2e^{2}A}{7\pi\gamma_{0}}(2\ln2)T,
\quad\sigma_{22,33}(T)=\frac{9e^{2}\upsilon^{2}}{20\pi\gamma_{0}}.
\end{align}
Here $\sigma_{22,33}(T)$ is temperature independent, which is usually the case of isotropic systems.
However, $\sigma_{11}(T)$ shows $T$ linear behavior.

In fact, as long as
the scattering rate does not show any momentum dependence, $\sigma_{11}(T)/\sigma_{33}(T)$
shows a universal behavior given by
\begin{align}
\frac{\sigma_{11}(T)}{\sigma_{33}(T)}=C_{0}\frac{A}{\upsilon^{2}}T,\quad
C_{0}=\frac{40}{63}\frac{\int dx (-\frac{df}{dx})|x|^{5/2}}{\int dx (-\frac{df}{dx})|x|^{3/2}}\approx 1.8,
\end{align}
where
\begin{align}
\sigma_{11}(T)&=\frac{2e^{2}\sqrt{A}}{7\pi^{2}\upsilon^{2}}\int d\varepsilon |\varepsilon|^{5/2}\Big(-\frac{\partial f}{\partial\varepsilon}\Big)\tau(\varepsilon)
\nonumber\\
\sigma_{33}(T)&=\frac{9e^{2}}{20\pi^{2}\sqrt{A}}\int d\varepsilon |\varepsilon|^{3/2}\Big(-\frac{\partial f}{\partial\varepsilon}\Big)\tau(\varepsilon).
\end{align}


\begin{thebibliography}{99}
\bibitem{DFT}
Details for first-principles calculations can be found in \cite{BiTeI_theory2}.
\bibitem{TPT}
The condition for the topological phase transition mediated by
a intervening semi-metallic phase can be found in \cite{Murakami}.
\bibitem{disorder_conductivity}
The $T$ linear behavior of $\sigma_{11}(T)$ due to disorder scattering implies
a vanishingly small conductivity
in $T\rightarrow0$ limit. This is an artifact of the simple (non self-consistent)
Born approxiamtion. Once the self-consistency is carefully considered,
$\sigma_{11}(T)$ should approach a constant value as $T\rightarrow0$.
\end{thebibliography}
\end{document}